\def\gsim{\;\lower4pt\hbox{${\buildrel\displaystyle >\over\sim}$}\,}
\def\lsim{\;\lower4pt\hbox{${\buildrel\displaystyle <\over\sim}$}\,}
\def\mach{{\cal M}}
\def\em{{\rm em}}
\def\ener{{\overline{\cal E}}\,}
\def\FLASH                 {{\sc flash}}
\def\PARAMESH              {{\sc paramesh}}

\newcommand{\referee}{ }

\documentclass{aa}

\usepackage{graphicx}
\usepackage{natbib}
\usepackage[english]{babel}
\usepackage{txfonts}
\begin{document}
   \title{Observability and diagnostics in the X-ray band of
          shock-cloud interactions in supernova remnants}

   \author{S. Orlando\inst{1},
           F. Bocchino\inst{1},
           M. Miceli\inst{2,1},
           X. Zhou\inst{3,1},
           F. Reale\inst{2,1},
          \and
           G. Peres\inst{2,1}
          }

   \offprints{S. Orlando,\\ e-mail: orlando@astropa.inaf.it}

   \institute{INAF - Osservatorio Astronomico di Palermo ``G.S.
              Vaiana'', Piazza del Parlamento 1, I-90134 Palermo, Italy
         \and
              Dip. di Scienze Fisiche \& Astronomiche, Univ. di
              Palermo, Piazza del Parlamento 1, I-90134 Palermo,
              Italy
         \and
              Department of Astronomy, Nanjing University, Nanjing
              210093, China
             }

   \date{Received \quad\quad\quad ; accepted \quad\quad\quad }

   \authorrunning{S. Orlando et al.}
   \titlerunning{Observability and diagnostics in the X-ray band of
                 shock-cloud interactions in SNRs}

% \abstract{}{}{}{}{}
% 5 {} token are mandatory

  \abstract
   {X-ray emitting features originating from the interaction of supernova
    shock waves with small interstellar gas clouds are revealed in many
    X-ray observations of evolved supernova remnants (e.g. Cygnus Loop
    and Vela), but their interpretation is not straightforward.}
   {We develop a self-consistent method for the analysis and interpretation
    of shock-cloud interactions in middle-aged supernova remnants, which
    can provide the key parameters of the system and the role of relevant
    physical effects like the thermal conduction, without the need to
    run ad-hoc numerical simulations and to bother of morphology details.}
   {We explore all the possible values of the shock speed and cloud
    density contrast relevant to middle-aged SNRs with a set of
    hydrodynamic simulations of shock-cloud interaction, including the
    effects of thermal conduction and radiative cooling. From the
    simulations, we synthesize spatially and spectrally resolved
    focal-plane data as they would be collected with XMM-Newton/EPIC,
    an X-ray instrument commonly used in these studies.}
   {We devise and tune up two diagnostic tools, the first based on
    the mean-photon energy vs. count rate scatter plot and the second on
    the spectral analysis of the interaction region, that can be used to
    highlight the effects of thermal conduction and to derive the shock
    speed in case of efficient conduction at work. These tools can be
    used to ascertain information from X-ray observations, without
    the need to develop detailed and ad-hoc numerical models for the
    interpretation of the data.}
   {}


   \maketitle

%
%________________________________________________________________
\section{Introduction}
\label{intro}

Supernova remnants (SNRs) are known to be a privileged laboratory
to investigate the physical and chemical evolution of the galactic
interstellar medium (ISM) and the mass distribution of the plasma in
the Galaxy. Multi-wavelength observations of evolved SNRs (e.g.,
\citealt{1995ApJ...444..787G}; \citealt{2000A&A...359..316B};
\citealt{2002AJ....124.2118P}; \citealt{2004ApJ...610..285N};
\citealt{2005A&A...442..513M}) can be a useful tool to investigate 
the physics of SNRs, for instance the interaction of the remnants with
inhomogeneities (clouds) of the ISM. However, this interaction
involves many non-linear physical processes (e.g. radiative losses and
thermal conduction) which make the analysis of the observations
quite difficult. A further limitation comes from the superposition of
different emitting regions along the line-of-sight (hereafter LoS) and,
in most cases, the data interpretation is not unique.

A powerful approach in the data analysis is based on hydrodynamic
and MHD simulations of the shock-cloud interaction, which takes into
account the most relevant physical mechanisms (e.g. thermal conduction,
radiative cooling, etc.), and on the comparison of the model results with
observations. Previous studies (\citealt{2005A&A...444..505O,orlando2,
2006A&A...458..213M, 2008ApJ...678..274O}) were devoted to investigate,
through numerical modeling, the interaction of SNR shock fronts with
small interstellar gas clouds. The scope includes: i) to investigate
the role of the different physical processes at work on the dynamics
and energetic of the shocked cloud, and ii) to analyze accurately the
SNRs observations through their comparison with model results.

As a part of this project, we already investigated the role of
thermal conduction and radiative cooling on the evolution of the
shocked cloud in the unmagnetized limit. We explored two physical
regimes in which each of the two physical processes in turn dominates
(\citealt{2005A&A...444..505O}, hereafter Paper I) and found that, in
general, the thermal conduction determines the evaporation of a fraction
of the shocked cloud, forming a hot and tenuous gas phase (the corona)
surrounding the cloud core. In the presence of an organized interstellar
magnetic field, the thermal conduction is known to be inhibited across
the magnetic field lines and the radiative cooling can be enhanced
due to magnetic plasma confinement. We explored the role played by the
magnetic-field-oriented thermal conduction and the radiative cooling
during the shock-cloud interaction, considering different configurations
of the magnetic field (\citealt{2008ApJ...678..274O}). We found that
the magnetized cases fall in between the limit of completely suppressed
thermal conduction and the unmagnetized limit with conduction.

Our numerical models have also been used to make predictions on the
expected X-ray emission from the shock-cloud interaction.  We showed
that the X-ray emitting structures do not trace the morphology of the
flow structures originating from the shock-cloud interaction and that
the shocked clouds are visible more easily during the early phases of
their evolution (\citealt{orlando2}, hereafter Paper II).

However, the big effort done in the modeling of shock-cloud
interaction and its X-ray emission has not been counterbalanced by
a rigorous methodology in the comparison between X-ray observations
of SNR shells and models. The high resolution instruments on board
XMM-Newton and Chandra have provided us with excellent images and
spectra of SNRs which are always much more complicated than the ideal
cases treated in numerical simulations. Therefore, a straightforward
comparison between models and observations is still difficult,
and this tends to hamper our understanding of the details of the
physical processes which are at the base of the X-ray radiation from
SNR shells. \citet{2006A&A...458..213M} made one of the first attempt in
filling the gap between models and observations. They compared the X-ray
observations of an isolated knot in the northern rim of the Vela SNR
(Vela FilD; \citealt{2005A&A...442..513M}) with an ad-hoc hydrodynamic
model; the comparison showed that the bulk of the X-ray emission in the
knot originates in the cloud material heated by the transmitted shock
front, but significant X-ray emission is also associated to the cloud
material which evaporates in the intercloud medium, under the effect of
the thermal conduction. While this strategy has proved to be winning, it
is quite model-dependent, in the sense that it is based on an accurate
and strict morphological and spectral comparison which may be time and
resource consuming.

On the contrary, the idea behind this paper is that the exploration of the
parameter space of the shock-cloud model already performed in Paper I and
Paper II, along with the extension presented here toward still unexplored
values of the cloud density contrast, may be used to devise a quick and
effective methodology for the interpretation of current generation X-ray
satellite observations of shock-cloud interactions, without the need of
running ad-hoc numerical models. Our intent is to provide easy-of-use
recipes that allow to extract from the data many of the key parameters
governing the evolution of shocked clouds, by comparison with a set of
model quantities normalized in a way to eliminate the dependence from
unnecessary details, like e.g. the exact morphology of the hit cloud. In
particular, our scope includes to devise a diagnostic tool able to quickly
assess if the spectral results obtained in the interaction regions are
dominated by thermal conduction, a physical effect whose contribution
to the X-ray emission is modulated by the magnetic field and, therefore,
still in debate and uncertain.

The paper is organized as follows: Sect. \ref{sec2} briefly describes
the numerical setup, the physical parameters of the problem, and the
method to synthesize, from the numerical simulations, X-ray observations
as they would be obtained with X-ray observatories; Sect. \ref{result}
presents the results of the numerical simulations; in Sect. \ref{diag}
we describe the diagnostic tools devised in this paper and apply the
methods, as an example, to X-ray observations reported in the literature;
in Sect. \ref{conc} we draw our conclusions.

%__________________________________________________________________
\section{Hydrodynamic modeling}
\label{sec2}

We model the three-dimensional interaction of a SNR shock front with an
ISM cloud in the same way we have done in Paper I, to which the reader
is referred to have more details. We summarize here the main model
features. The cloud is assumed to be small compared to the curvature
radius of the shock\footnote{This assumption is valid for a 1 pc cloud
in the middle-aged SNRs Vela and Cygnus Loop, whose shell has a radius $>
10$ pc.} and in pressure equilibrium with the unperturbed isothermal and
homogeneous ambient medium; we consider, therefore, a planar shock front
and an isobaric cloud, spherical for simplicity. The shock propagates with
a Mach number $\mach \gg 1$ in the ambient medium. The post-shock initial
conditions are given by the strong shock limit (\citealt{zel66}). The
fluid is assumed fully ionized, and is regarded as a perfect gas.

The plasma dynamics is described by solving numerically the
time-dependent fluid equations of mass, momentum, and energy
conservation (see Eqs. 1-5 in Paper I). The model takes into account
the thermal conduction (\citealt{spi62}) and the radiative losses from
an optically thin plasma (e.g. \citealt{rs77}, \citealt{mgv85}, and
\citealt{2000adnx.conf..161K}). The thermal conduction includes the
free-streaming limit (saturation) on the heat flux (\citealt{cm77},
\citealt{1984ApJ...277..605G}, \citealt{1989ApJ...336..979B},
\citealt{2002A&A...392..735F}, and references therein). Our calculations
also include a passive tracer associated with the cloud material to
trace its motion during the evolution. \referee{A discussion of the
assumptions of the model and their influence on the results is presented
in Sect.~\ref{limits}.}

The numerical code is \FLASH\ (\citealt{for00}), a multidimensional
hydrodynamics code for simulating astrophysical plasmas, which uses the
\PARAMESH\ (\citealt{mom00}) library for block-structured adaptive mesh
refinement (AMR), and has been customized with numerical modules that
treat thermal conduction and optically thin radiative losses (see Paper
I for details). The initial configuration, the boundary conditions,
and the AMR setup of the simulations used here are the same as those
adopted and discussed in Paper I.

We consider, as a reference case, the $\mach = 50$ shock model
described in Paper I; we then explore the parameter space by varying,
alternatively, either the Mach number, $\mach$, or the density contrast
cloud/surrounding medium, $\chi$. In the reference model (RCm50c10),
the unperturbed ambient medium is at temperature $T_{\rm ism} = 10^4$ K
and particle number density $n_{\rm ism} = 0.1$ cm$^{-3}$, the spherical
isobaric cloud has a radius $r_{\rm cl} = 1$ pc and density contrast
$\chi = 10$ (particle number density $n_{\rm cl} = \chi n_{\rm ism} = 1$
cm$^{-3}$). The SNR shock front is planar at Mach number $\mach = 50$
and temperature $T_{\rm psh} = 4.7$ MK. In the other simulations, the
Mach number varies in the range $40\leq \mach \leq 60$ (corresponding
to shock temperatures in the range $3$ MK $\leq T_{\rm psh} \leq 7$
MK) and the cloud density contrast in the range $3\leq \chi \leq 30$
(corresponding to particle number density of the cloud in the range $0.3
\mbox{ cm$^{-3}$} \leq n_{\rm cl} \leq 3 \mbox{ cm$^{-3}$}$). We note that
in Paper II, we have already partially explored the variation induced by
a different choice of the shock speed (we have considered $\mach = 30$
and 50 at $\chi = 10$). Here, we present for the first time the results
for different density contrasts. These ranges are representative of most
of the shock-cloud interaction regions observed in evolved SNRs (e.g. 
Vela, Cygnus Loop, and G296.5+10.0).

\begin{table}[!t]
 \caption{Parameters of the simulated shock-cloud interactions.}
 \label{tab1}
 \begin{center}
 \begin{tabular}{lccccccc}
 \hline
 \hline
 Run & $\mach^a$ & $\chi^b$ & $w^c$ & $T_{\rm psh}^d$
 & $\tau_{\rm cc}^e$ & therm. \\
 &  & & [km s$^{-1}$] & [MK] & [$10^3$ yr] & cond.\\
 \hline
 HYm40c10   & 40 & 10 & 458 & 3.0 & 6.75 & no  \\
 HYm50c10   & 50 & 10 & 574 & 4.7 & 5.41 & no  \\
 HYm60c10   & 60 & 10 & 688 & 6.7 & 4.50 & no  \\
 RCm40c10   & 40 & 10 & 458 & 3.0 & 6.75 & yes \\
 RCm50c10   & 50 & 10 & 574 & 4.7 & 5.41 & yes \\
 RCm60c10   & 60 & 10 & 688 & 6.7 & 4.50 & yes \\
 RCm50c03   & 50 & 03 & 574 & 4.7 & 2.96 & yes \\
 RCm50c30   & 50 & 30 & 574 & 4.7 & 9.37 & yes \\
 \hline
 \\
 \end{tabular}\\
 \end{center}
$^a$\,Shock Mach number. $^b$\,Density contrast cloud/ambient medium.
$^c$\,Velocity of the SNR shock. $^d$\,Temperature of the post-shock ambient
medium. $^e$\,Cloud crushing time (\citealt{1994ApJ...420..213K}).
 \end{table}

\begin{figure}[!t]
  \centering
  \includegraphics[width=8.cm]{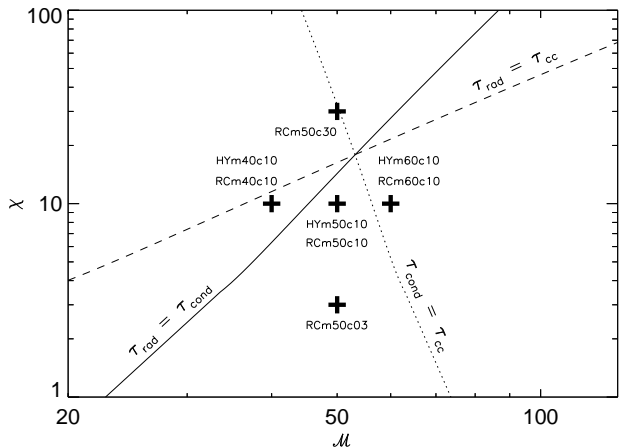}
  \caption{$\chi-\mach$ parameter space (adapted form Paper I). The
     lines are derived for length-scale $l = 1$ pc, and for an unperturbed
     ambient medium with temperature $T_{\rm ism} = 10^4$ K and particle
     number density $n_{\rm ism} = 0.1$ cm$^{-3}$ (see Paper I for
     details): the solid line separates regions dominated by radiative
     cooling (on the left) from regions dominated by thermal conduction
     (on the right); the dashed line marks the density contrast, $\chi$,
     above which the cooling time-scale $\tau_{\rm rad}$ is shorter
     than the cloud crushing time $\tau_{\rm cc}$; the dotted line marks
     the values of $\chi$ above which the thermal conduction time-scale
     $\tau_{\rm cond}$ is shorter than $\tau_{\rm cc}$. The parameter
     pairs explored are marked with crosses.}
\label{fig1} \end{figure}

The effects of thermal conduction on the shocked cloud evolution have
been fully investigated in Papers I and II. Since we are interested
in deriving some diagnostic to be used in real X-ray observations,
we have compared runs with this physical process (hereafter RC runs)
with other runs without it (hereafter HY runs). As shown in a previous
work (\citealt{2008ApJ...678..274O}), shock-cloud interactions in an
organized interstellar magnetic field fall in between these two limits
(i.e. HY and RC cases). A summary of all the simulations discussed
in this paper is in Table \ref{tab1}, while Fig. \ref{fig1} shows
the simulations in the $\chi-\mach$ parameter space. As discussed in
Paper I (cfr. Fig. 2 in Paper I), this plot can be used to evaluate
if radiative cooling is competitive with respect to thermal conduction
for a given run. For example, the shock transmitted into the cloud is
strongly radiative in runs RCm40c10 and RCm50c30. On the other hand, the
thermal conduction dominates over the radiative losses in all the other
cases (i.e. RCm50c03, RCm50c10 and RCm60c10): the cloud is expected to
evaporate on a time-scale comparable (in RCm50c03 and RCm50c10) or shorter
(in RCm60c10) than $\tau_{\rm cc}$.

\subsection{Synthesis of the X-ray observations}
\label{emvst-spec}

The output of the numerical simulations is the evolution of temperature,
density, and velocity of the plasma in the spatial domain. From the
density and temperature values, we synthesize spatially and spectrally
resolved X-ray observations with the XMM-Newton/EPIC-pn X-ray imaging
spectrometers (\citealt{sbd01}). The method can be easily extended to
other X-ray instruments.

The emission measure in the $j$th domain cell is calculated as $\em_{\rm
j} = n_{\rm Hj}^2 V_{\rm j}$, where $n_{\rm Hj}^2$ is the particle
number density in the cell, and $V_{\rm j}$ is the cell volume. We
assume that the direction of the LoS corresponds to the $y$ axis (in
the Cartesian coordinate system), perpendicular to the direction of
propagation of the SNR shock front, and that the depth along the LoS
is 10 pc (a typical value for the shell of evolved SNRs, such as in
Vela and in Cygnus Loop). We then derive the distributions of
emission measure versus temperature, EM($T$), integrated along the LoS
for each $(x,z)$, in the temperature range $5 < \log T\mbox{(K)}< 7$
(divided into 50 bins, all equal on a logarithmic scale). From the
EM($T$) distributions, we synthesize maps of X-ray emission and X-ray
spectra, using the MEKAL spectral synthesis code (\citealt{mgv85};
\citealt{kaas92}, \citealt{2000adnx.conf..161K}), assuming solar metal
abundances (\citealt{1991sia..book.1227G}).

We assume the source to be at a distance $D_{\rm snr} = 500$ pc (as, for
instance, in the case of Cygnus Loop) and we filter the spectra
through an ISM absorption column density, $N_{\rm H} = 5\times10^{20}$
cm$^{-2}$ (\citealt{mm83}), according to typical values derived from
SNR observations at that distance (e.g. \citealt{2002AJ....124.2118P}).
The absorbed X-ray spectra are then folded through the instrumental
response to obtain focal plane spectra. The exposure time is assumed to
be $t_{\rm exp} = 10$ ks for EPIC-pn (see, for instance,
\citealt{2005A&A...442..513M}). The photon counts are randomized in each
energy instrumental channel of the focal-plane spectra according to
Poisson statistics, using the rejection method applied to the Poisson
distribution (\citealt{press86}). X-ray emission maps are produced in
selected energy bands, assuming a spatial resolution of 4 arcsec; the
X-ray images are convolved with the corresponding point spread function
(PSF), as given by \citet{ghizza02} for EPIC-pn.

The final products are X-ray simulated observations, spatially and
spectrally resolved, in a format virtually identical to that of real
observations collected with EPIC-pn. To such data, we apply the standard
methods of analysis commonly used for X-ray observations.

%__________________________________________________________________
\section{Results}
\label{result}

\subsection{Light-curves}

\begin{figure}[!t]
  \centering
  \includegraphics[width=7.8cm]{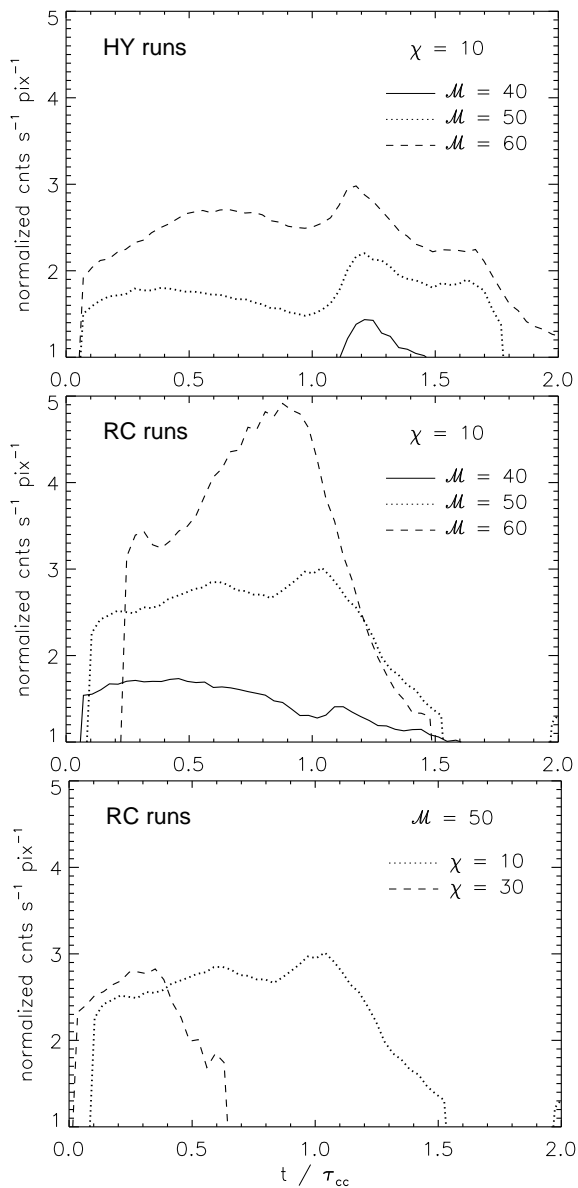}
  \caption{X-ray light-curves of the shock-cloud interaction region in
           the $[0.3-2]$ keV band (see text for details on the
           definition). The time dependent surface brightness is
           normalized to the average value of the post-shock intercloud
           region. Upper panel: runs without thermal conduction (HY
           runs) with $\chi=10$ and $\mach=40,50,60$. Middle panel:
           runs with thermal conduction (RC runs) with $\chi=10$
           and $\mach=40,50,60$. Lower panel: RC runs with $\mach=50$
           and $\chi=10, 30$.}
\label{fig2}
\end{figure}

The detectability of the shock-cloud collision in the X-ray band is
expected to depend on the $\mach$ and $\chi$ parameters. From our
simulations, we derive the X-ray light curves of the region associated
to the shocked cloud, to understand at which stage of the interaction
the visibility of the cloud is maximum. Such a region is selected in
each synthesized EPIC-pn count rate image in the $[0.3-2.0]$ keV
band (typically selected for the analysis of evolved SNR shock-cloud
interaction; see, for instance, \citealt{2005A&A...442..513M}), by
considering all the pixels having a median energy of X-ray photons,
$\ener$ (\citealt{2004ApJ...614..508H}), which is less than 90\% of the
$\ener$ derived for the surrounding medium\footnote{In fact, the shocked
cloud material is expected to be cooler than the surrounding medium.};
from these pixels, then, we evaluate the average counts s$^{-1}$ per
pixel, $F_{\rm X}$, normalized to the value derived for the intercloud
medium. The X-ray light curves reported in Fig. \ref{fig2} show when the
shocked cloud is detectable (when the normalized $F_{\rm X}$ is higher
than 1) and its X-ray luminosity is maximum.

In general, we find that the higher is $\mach$, the higher the normalized
$F_{\rm X}$ at each stage of the evolution (see Fig.~\ref{fig2}). In most
cases, the shocked cloud is visible in the X-ray band in the interval
$0.1~ \tau_{\rm cc} < t < 1.5~\tau_{\rm cc}$. The thermal conduction makes
the shocked cloud brighter than in cases without conduction, broadening
the peak in the X-ray light-curve for any Mach number. In fact, the
conduction contributes to the cloud heating, increasing the amount of
cloud material above 1 MK and emitting in the X-ray band. Note also that
the conduction makes the shocked cloud hardly detectable in cases with
$\chi \leq 3$ due to the quick evaporation of the cloud. \referee{For
instance, in run RCm50c03, there are no pixels with $\ener$ below 90\%
of the value derived for the background and no light-curve
can be reported in Fig.~\ref{fig2}.} On the other hand, the conduction
makes shocked clouds with $\chi \gsim 20$ (whose evolution is strongly
dominated by the radiative losses) partially visible during the very
early phases of the evolution (see lower panel in Fig.~\ref{fig2});
these clouds would be not detectable in X-rays, in the absence of thermal
conduction, being their temperature $T_{\rm scl}\approx 2.5\times 10^5$
K. In these cases, only the thermally conducting corona is detected,
being the core much cooler than 1 MK.

\subsection{Spectral analysis}
\label{spec_ana}

\begin{figure}[!t]
  \centering
  \includegraphics[width=8.5cm]{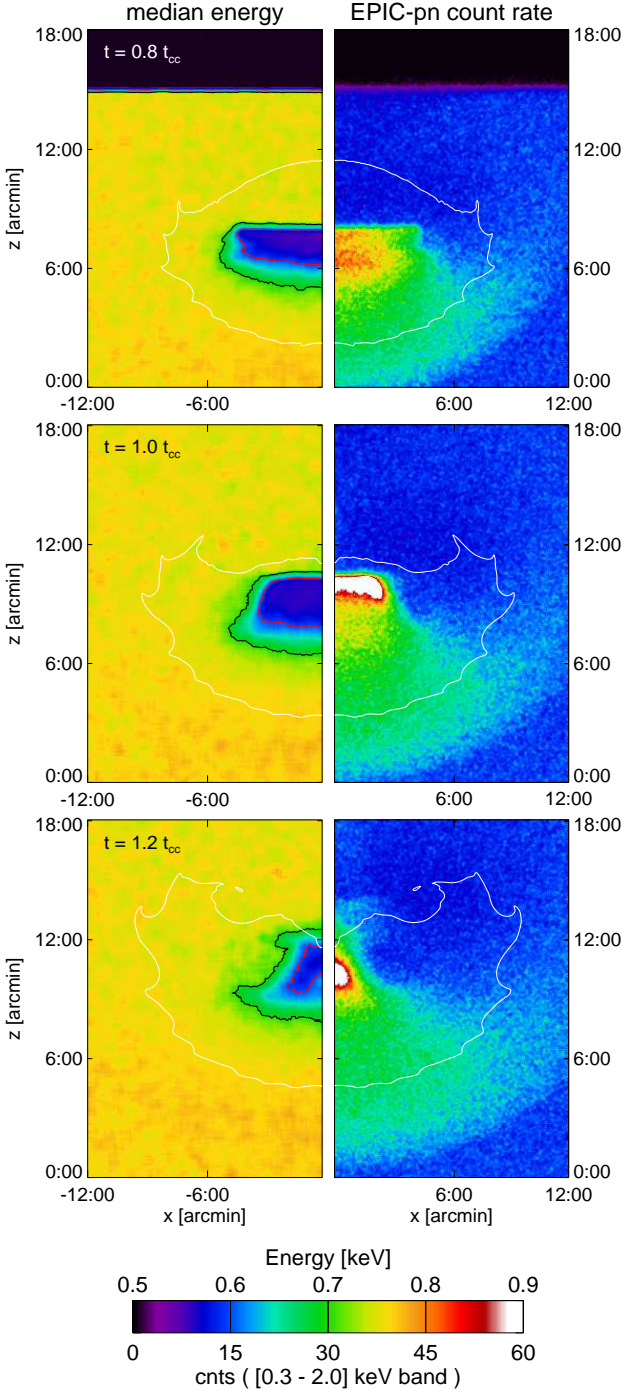}
  \caption{Median energy maps (left) and EPIC-pn count rate images
   (right) in the [$0.3-2.0$] keV band derived for run RCm50c10 at the
   three labeled times during the evolution. The pixel size is $\sim 4$
   arcsec and the exposure time is 10 ks. The images are smoothed with
   a boxcar of width $\sigma = 12$ arcsec. The white contours mark the
   cross-section of the cloud on the plane of the image, identified by
   zones consisting of the original cloud material by more than 90\%;
   the contours superimposed to the median energy maps mark the bright
   knot (red) and the diffuse region (diffuse region, DR; black).}
\label{fig3} \end{figure}

The shocked cloud is detectable with EPIC-pn during the early
phases of the shock-cloud interaction ($t < 1.5 ~\tau_{\rm cc}$; see
Fig.~\ref{fig2}) as a bright knot surrounded by a diffuse region (see
right panels in Fig.~\ref{fig3}). We focus the spectral analysis in the
interval $0.4~ \tau_{\rm cc} \leq t \leq 1.4~ \tau_{\rm cc}$ when the
shocked cloud is the brightest in all the models. For each sampled
X-ray image, we select spatial sub-regions in the computational domain
and analyze the X-ray spectra extracted from each of them. To select
spectrally homogeneous regions, we derive maps of the median energy
of X-ray photons, $\ener$, from the EPIC-pn data that allows to convey
at the same time both spatial and spectral information on the emitting
plasma (\citealt{2004ApJ...614..508H}). Since the shocked cloud is cooler
than the surrounding medium, the median photon energy, $\ener$, of the
bright region is lower than that of the surroundings (see left panels
in Fig.~\ref{fig3}). Thus we select subregions with a median photon
energy $0.5 \mbox{ keV} < \ener < 0.6 \mbox{ keV}$ (to identify the
knot), and subregions with $0.6 \mbox{ keV} < \ener < 0.7 \mbox{ keV}$
(for the diffuse region, DR). The knot corresponds to the brightest
portion of the X-ray image in Fig. \ref{fig3}, and the DR selects the
intermediate brightness region surrounding the knot (compare left and
right panels in Fig.~\ref{fig3}).

It is important to stress that this definition of the extraction regions
for spectral analysis are completely independent from the morphology
of the X-ray emission. This has the great advantage that it can be
straightforwardly applied to any current X-ray telescope observation for
which the mean photon energy map can be computed. Moreover, it makes the
spectral analysis independent, in first approximation, from the details
of the shape of the ISM clouds, which in reality may be much more complex
than the ideal spherical cloud proposed in our model.

The extracted spectra have a total number of photons ranging between
$10^4$ and $4\times 10^5$, adequate for a detailed spectral analysis. The
focal plane spectra have been analyzed using the spectral fitting
package XSPEC (\citealt{arn96}) and applying a multi-temperature fit
to each spectrum. All the extracted spectra are well fitted with two
MEKAL components of an optically-thin thermal plasma in collisional
ionization equilibrium (\citealt{mgv85}, \citealt{2000adnx.conf..161K}),
with solar abundances, and filtered through the interstellar absorption
(\citealt{mm83}). We have applied this procedure to any model listed in
Table \ref{tab1}, and in the interval $0.4~ \tau_{\rm cc} \leq t \leq 1.4~
\tau_{\rm cc}$ in step of $\delta t = 0.1~\tau_{\rm cc}$. We present here,
as an example, the results obtained in the reference models HYm50c10
and RCm50c10.

\begin{figure}[!t]
  \centering
  \includegraphics[width=9.cm]{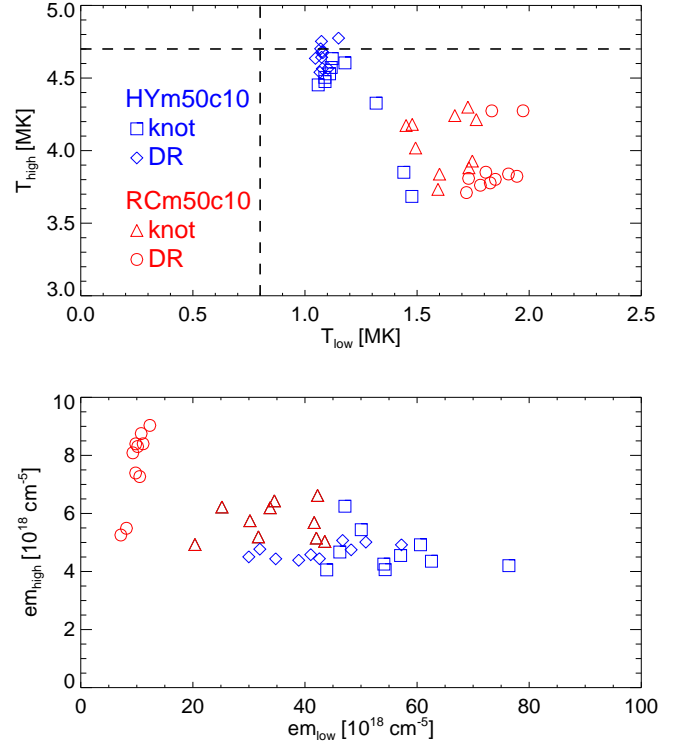}
  \caption{Best-fit values of temperature (upper panel) and emission
   measure per unit area (lower panel) for the EPIC-pn spectra extracted
   from the bright knot and from the diffuse region (DR) in runs
   HYm50c10 (blue) and RCm50c10 (red) at different epochs between 0.4
   and 1.4 $\tau_{\rm cc}$. The dashed lines in the upper panel mark
   the temperatures expected for the shock transmitted into the cloud
   ($T_{\rm scl}\approx 0.8$ MK) and for the shocked ambient plasma
   ($T_{\rm psh}\approx 4.7$ MK).}
  \label{fig4}
\end{figure}

Figure \ref{fig4} shows the temperature, $T$, and the emission measure
per unit area, $\em = \mbox{EM}/A_{\rm reg}$ (where $A_{\rm reg}$
is the area of the selected region), of the isothermal components
fitting the EPIC-pn spectra. When the thermal conduction is completely
suppressed (run HYm50c10), the spectra of both the knot and the DR
at the different epochs are described, in general, by two isothermal
components with temperatures $T_{\rm low}\approx 1$ MK and $T_{\rm high}
\approx 4.5$ MK; the emission measure of the hot component is $\em_{\rm
high} \approx 5\times 10^{18}$ cm$^{-5}$ in all the spectra, whereas
$\em_{\rm low}$ ranges between $3\times 10^{19}$ and $8\times 10^{19}$
cm$^{-5}$. Note that the temperature of the hot component is close
to the temperature of the shocked ambient plasma $T_{\rm psh} \approx
4.7$ MK, whereas $T_{\rm low}$ is slightly higher than the temperature
of the shock transmitted into the cloud, $T_{\rm scl} \approx 0.8$~MK
(see Paper I).

In run RCm50c10, the spectra are again described by two isothermal
components, but with some differences due to the thermal conduction. In
particular, $T_{\rm low}$ is higher and $T_{\rm high}$ is lower than the
values derived in HYm50c10, being the difference larger for the DR than
for the knot. Also $\em_{\rm low}$ is systematically lower and $\em_{\rm
high}$ higher than the values derived in HYm50c10. In general, we find
that the cold component is the most sensitive to the thermal conduction,
showing the largest differences between HYm50c10 and RCm50c10. The origin
of these differences is explained by comparing the results of the spectral
fitting with the distributions of emission measure per unit area versus
temperature, $\em(T)$, from which the extracted spectra originate.

\begin{figure*}[!t]
  \sidecaption
  \includegraphics[width=12.cm]{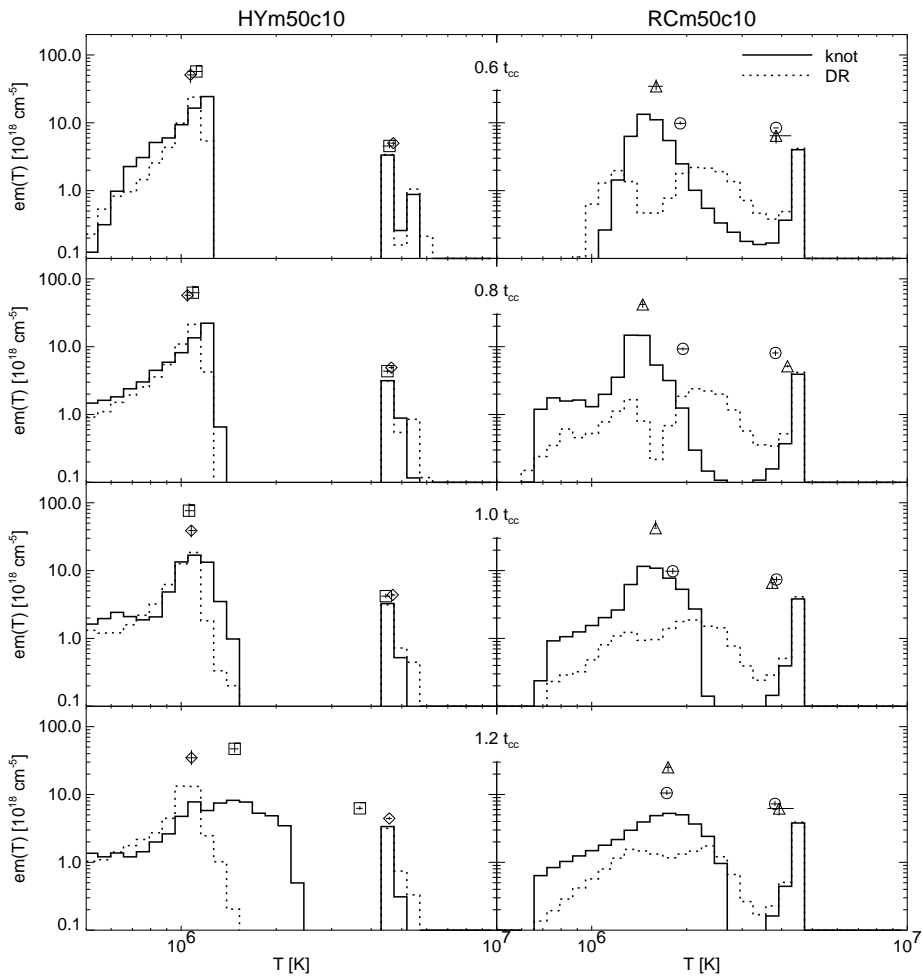}
  \caption{Distributions of emission measure per unit area versus
   temperature, $\em(T)$, in the range of temperature $5 < \log T
   (\mbox{ K}) < 7$ for the knot (solid) and for the diffuse region (DR,
   dotted) at the labeled times during the evolution. The figure shows the
   $\em(T)$ derived from run HYm50c10 (on the left) and from RCm50c10 (on
   the right). The figure also shows the results of the spectral fitting
   to the EPIC-pn spectra extracted from the selected regions (symbols
   as in Fig. \ref{fig4}). The errors are at 90\% confidence level.}
  \label{fig5}
\end{figure*}

Figure \ref{fig5} shows these $\em(T)$ distributions together with the
results of the spectral fitting. In general, the distributions of both
the knot and the DR are bi-modal. The cold peak around $T\approx 2$ MK
is due to the shocked cloud gas; the hot peak at $T\approx 5$ MK is due
to the shocked ambient plasma surrounding the cloud. The best-fit values
are localized around these maxima and, therefore, can be associated to
the shocked cloud gas (cold component) and to the shocked ambient plasma
(hot component).

In HYm50c10, the $\em(T)$ distributions of both the knot and the DR do not
change significantly during the evolution (except at $t\sim 1.2\;\tau_{\rm
cc}$, when the shocks transmitted from the front and from the rear of
the cloud interact; see Paper II), with the two bumps steadily centered
around the temperatures expected for the shock transmitted into the cloud
($T_{\rm scl}\approx 0.8$ MK) and for the shocked ambient plasma ($T_{\rm
psh}\approx 4.7$ MK), respectively. By comparing HYm50c10 with RCm50c10,
the main effects of the thermal conduction are: i) to smooth the $\em(T)$
distributions, because of a transition region formed between the inner
part of the cloud and the ambient medium in which the density decreases
and the temperature increases smoothly in the radial direction (see Paper
II); and ii) to shift the first bump to higher temperatures due to the
gradual thermalization of the shocked cloud material to the temperature
of the shocked ambient plasma (see also Paper II).

As a result of the conduction effects, the amount of plasma above 1
MK increases in RCm50c10, making the shocked cloud brighter in the
X-ray band (see Fig.~\ref{fig2}). Also, the changes in the $\em(T)$
distributions due to the thermal conduction determine the differences
in the results of the spectral fitting for HYm50c10 and RCm50c10 (see
Fig.~\ref{fig4}); for instance, the shift of the first bump in $\em(T)$
to higher temperatures leads to higher $T_{\rm low}$, and the smoothing
of $\em(T)$ leads to lower $\em_{\rm low}$ in RCm50c10. Note also that
the effects of the thermal conduction are the largest in the $\em(T)$
distribution of the DR, being the plasma of the corona surrounding the
cloud core subject to efficient heat conduction. As a consequence, the
cold fitting component in the DR is, in general, hotter than that in the
knot (see also Fig.~\ref{fig4}), whereas the opposite is true in HYm50c10
(i.e. the temperature of the shocked cloud material is never higher than
the temperature of the shock transmitted into the cloud).

\section{Diagnostics}
\label{diag}

\subsection{Median energy vs. count-rate scatter plot}

\begin{figure}[!t]
  \centering
  \includegraphics[width=7.7cm]{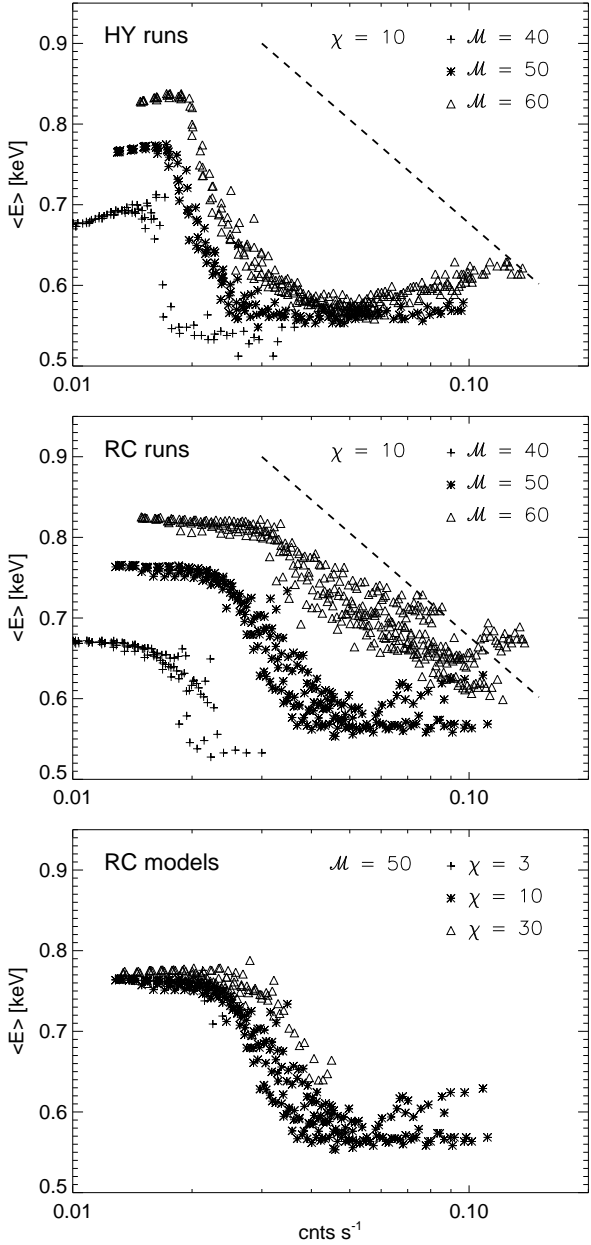}
  \caption{Median photon energy, $\ener$, versus count rate
           scatter plot derived from the EPIC-pn data in the period $0.4~
           \tau_{\rm cc} \leq t \leq 1.4~ \tau_{\rm cc}$. Upper panel:
           runs without thermal conduction (HY runs) with $\chi=10$ and
           $\mach=40,50,60$. Middle panel: runs with thermal conduction
           (RC runs) with $\chi=10$ and $\mach=40,50,60$. Lower panel:
           RC runs with $\mach=50$ and $\chi=3, 10, 30$. The dashed lines
           in the upper and middle panels show the slope of the $\ener$,
           versus count rate scatter plot derived from the analysis of
           EPIC data of Vela FilD (\citealt{2005A&A...442..513M}).}
\label{fig6}
\end{figure}

Since the thermal conduction modifies the temperature and density
structure of the shocked cloud (see Paper I), its effects may be expected
in the comparison of $\ener$ maps (related to the spatial distribution of
temperature) with count rate maps (related to the spatial distribution
of mass density). We derive, therefore, $\ener$ versus count rate
scatter plots (see, for instance, \citealt{2005A&A...442..513M}): we
first divide the range of count rate $[0.01-0.20]$ cnts s$^{-1}$ into
100 bins (all equal on linear scale); then, from the EPIC-pn count rate
images in the $[0.3-2.0]$ keV band, we derive the median photon energy
of all the pixels belonging to the same count rate bin. Fig. \ref{fig6}
shows the scatter plots derived for HY (upper panel) and RC (middle
and lower panels) runs at selected epochs between $0.4~ \tau_{\rm cc}
\leq t \leq 1.4~ \tau_{\rm cc}$ (when the shocked clouds are visible;
see Fig. \ref{fig2}). All these plots are characterized by a clear
descending trend and, then, in most of the cases, by a much flatter
fall (cold plateau): the higher the count rate, the lower is the median
energy and, therefore, the lower is the average temperature along the
LoS. The descending branch and the cold plateau are the signature of the
shock-cloud collision: the former roughly corresponds to the DR and the
latter to the bright and cold knot defined in Sect. \ref{spec_ana}.

In HY runs, scatter plots derived for $\chi=10$ and different $\mach$
show a similar shape, characterized by a very steep descending branch and
a well defined cold plateau (see upper panel in Fig. \ref{fig6}). The
descending branch shows an abrupt transition between the intercloud
material (highest $\ener$) and the shocked cloud material (lowest
$\ener$). In RC runs (see middle panel in Fig. \ref{fig6}), the thermal
conduction makes the slope of the descending branch flatter than that
of HY runs, being the absolute value of the slope smaller for higher
$\mach$. The flattening of the descending branch reflects a smooth
temperature and density structure of the shocked cloud due to the heat
conduction that leads to the gradual growth of a transition region from
the inner part of the cloud to the ambient medium (see Paper I). Note
that the scatter plots of runs with different $\chi$ and same $\mach=50$
are virtually indistinguishable (see lower panel in Fig. \ref{fig6})
independently of the role of radiation or conduction. This is due to the
fact that the contribution to X-ray emission invariably comes from regions
dominated by thermal conduction. In particular, in run RCm50c30, the
X-ray emission originates from the thermally conducting corona, being
the cloud core at temperatures $T \lsim 2\times 10^5$ K.

In summary, these scatter plots can be very useful to determine the role
of the thermal conduction through the slope of the descending branch
and to derive hints about the speed of the shock (in case of efficient
conduction at work). On the other hand, the plots are poorly useful to
infer the actual density contrast of the cloud.

\subsection{Temperature and emission measure ratios}

\begin{figure}[!t]
\centering
\includegraphics[width=8.5cm]{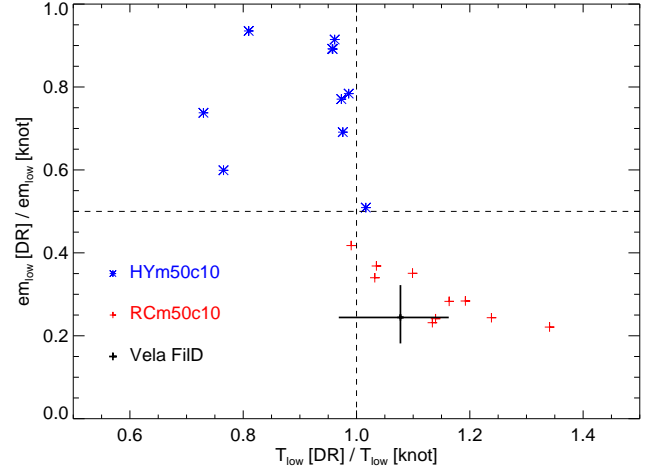}
\caption{The figure compares the temperature and the emission measure
 per unit area of the cold component derived for the knot with those
 derived for the diffuse region (DR). The blue stars (red crosses)
 mark the values derived in run HYm50c10 (RCm50c10) and reported in
 Fig.~\ref{fig4}; the black symbol with error bars mark the values
 derived from the analysis of EPIC data of Vela FilD (regions 4, knot,
 and 1, diffuse region, in \citealt{2005A&A...442..513M}).}
\label{fig7} \end{figure}

The spectral analysis discussed in Sect.~\ref{spec_ana} suggests that
the cold fitting components describing the knot and the DR are sensitive
to the thermal conduction; we propose, therefore, to use them as a
diagnostic tool to trace the efficiency of conduction. Fig. \ref{fig7}
compares the temperature and emission measure values derived for the
knot with those derived for the DR (and reported in Fig.~\ref{fig4}) in
runs HYm50c10 and RCm50c10. The runs with/without thermal conduction
are clearly separated in the plot, the thermal conductive case being
localized in the bottom-right quadrant and the pure hydrodynamic case in
the top-left quadrant. This result is determined by the development of
the thermally conducting corona in RCm50c10 and, therefore, is expected
to be general. In particular, as discussed in Sect.~\ref{spec_ana}, the
thermal conduction smooths the first bump in the $\em(T)$ distributions
and shifts it to higher temperatures, being this effect larger for the DR
than for the knot. As a result, the cold fitting component derived for the
DR is, in general, hotter than that derived for the knot (i.e. $T_{\rm
low}[{\rm DR}] > T_{\rm low}[{\rm knot}]$), whereas the opposite is
true (i.e. $T_{\rm low}[{\rm DR}] < T_{\rm low}[{\rm knot}]$) if the
conduction is suppressed. Also, the smoothing of the $\em(T)$ distribution
due to the conduction is the largest for the DR (because the corona
surrounding the cloud core is subject to efficient heat conduction),
leading to the smallest values of $\em_{\rm low}$ (see also lower panel in
Fig.~\ref{fig4}). 

In summary, the temperature and emission measure ratios are an excellent
way to determine the role of the thermal conduction in the evolution of
the system.

\subsection{An example of model vs. observation comparison}
\label{comp}

Our study shows that evidence of thermal conduction at work during the
shock-cloud interaction may be found in the spectral analysis of X-ray
data. As discussed in the previous sections, the $\ener$ vs. count-rate
scatter plot and the temperature and emission measure ratios can be
efficient diagnostic tools to derive the shock speed and the role of
thermal conduction, which, in turn, is linked to the magnetic field
configuration, as shown by \cite{2008ApJ...678..274O}. In this section,
we challenge the above diagnostic tools and show that they can be easily
used in the analysis of X-ray data, by comparing our model results with
X-ray observations reported in the literature.

In particular, we focus on a well-studied region, the ``FilD
region", that is an isolated, bright X-ray knot in the northern
rim of the Vela SNR. Because of its proximity ($\sim 250$ pc,
\citealt{1999A&A...342..839B}, \citealt{1999ApJ...515L..25C}) Vela
is an ideal target for this kind of study, allowing us to observe the
interaction of the SNR shock front with relatively small clouds, like
FilD ($\sim 2\times10^{18}$ cm; see \citealt{2005A&A...442..513M})
in great detail. The analysis of an XMM-Newton observation of FilD
(\citealt{2005A&A...442..513M}) has shown that its X-ray spectra can
be modeled by an optically-thin plasma with two thermal components (at
$\sim 1$ MK and $\sim 3$ MK, respectively) with inhomogeneous volume
distributions along the line of sight. The cold component dominates in
the brightest region that is surrounded by a diffuse region with harder
X-ray emission. To interpret these results, \citet{2006A&A...458..213M}
developed a detailed hydrodynamic model of FilD, synthesized X-ray
emission maps and spectra from the model, and compared them with the
data. Their analysis has shown that the X-ray and optical emission of
FilD can be explained as the result of the interaction of a SNR shock
(with Mach number $\mach=57$) with an ellipsoidal cloud 30 times denser
than the intercloud medium; the estimated interaction time is $\sim
0.32~\tau_{\rm cc}$. \citet{2006A&A...458..213M} proved that the two
components originate in the cloud material heated by the transmitted
shock front and by heat conduction between the cloud and the hotter,
shocked intercloud medium. FilD, therefore, is an ideal benchmark
for our model, being a case for which the thermal conduction has been
proved to be at work. Since the parameters used in our simulations are
slightly different from the parameters deduced from the observations
(including the shape of the cloud), we do not expect a perfect match,
but the comparison will nonetheless give us many useful information.

Among the runs presented here, the one matching the density contrast
of the shock-cloud interaction is RCm50c30 (see Tab.~\ref{tab1}). As
already discussed, the shocked cloud with such a density contrast ($\chi
\sim 30$) would not be detectable in X-rays if the thermal conduction is
suppressed, being its estimated temperature $T_{\rm scl}\approx 3.5\times
10^5$ K. On the other hand, inspecting Fig.~\ref{fig2}, we note that,
in run RCm50c30 (dashed line in the lower panel), the shocked cloud is
visible in X-rays for a short time interval ($0.1-0.6 ~\tau_{\rm cc}$)
and has the largest surface brightness at the evolutionary stage estimated
for FilD ($\sim 0.32~\tau_{\rm cc}$). We expect a brighter emission for
higher values of $\mach$ (see middle panel in Fig.~\ref{fig2}). As shown
by our simulations, the detected X-ray emission originates in the cloud
material dominated by thermal conduction, confirming the relevance of
conduction in the evolution of FilD.

\citet{2005A&A...442..513M} analyzed the spectra extracted from the knot
and the DR composing the FilD region. Thus, we can derive the temperature
and emission measure ratios of the cold components derived by these
authors and plot them in Fig. \ref{fig7}. The observed values lie in
the bottom-right quadrant of the figure, confirming once again that in
FilD the thermal conduction is efficient, in perfect agreement with the
independent conclusion of \citet{2006A&A...458..213M}. The diagnostics
in Fig.~\ref{fig7} is of easy implementation and we suggest it as a
standard to check for the role of conduction.

\citet{2005A&A...442..513M} derived also a $\ener$ versus count rate
scatter plot for FilD that can be compared directly with the corresponding
scatter plots derived with our models. We overplotted a best-fit power-law
model (with index $= -0.25$) to the data of the FilD region reported in
Fig. 4 of \citet{2005A&A...442..513M}, considering the count rate as a
free parameter, a reasonable choice if we consider that the actual value
of the count rate depends on the actual LoS extension, which is poorly
known. The slope of the observed scatter plot is rather flat and cannot
be reproduced by models without the thermal conduction (see upper panel
in Fig.~\ref{fig6}). On the other hand, the observed slope is reproduced
quite well by our RC runs, in agreement with the evidence that the thermal
conduction plays an important role in the evolution of FilD. $\mach =
60$ seems to be the model which best reproduces the slope, in very
good agreement with the value obtained by \citet{2005A&A...442..513M}
($\mach = 57$) with a detailed analysis.

\subsection{Limits of the model}
\label{limits}

{In our simulations, we parametrize the thermal conductivity using the
classical Spitzer's conductivity and the saturation limit, assuming
essentially laminar thermal conduction in the all spatial domain. However,
regions of strong turbulence of different strength and extent can
develop in the system (especially in shock-cloud interactions dominated
by radiative cooling), for instance at the shear layers along the cloud
boundary or at the vortex sheets in the cloud wake. The turbulence in
these regions may have a significant effect on the thermal conduction,
leading to significant deviations of thermal conductivity from its
laminar values (e.g. \citealt{1983hppv.book.....G, 2001ApJ...562L.129N,
2006ApJ...645L..25L}). As a result, the thermal conduction may be
inhomogeneous due to the presence of turbulence. On the other hand,
the deviations of thermal conductivity from its laminar values are
expected to be relevant in the shocked intercloud medium, thus not
affecting our main conclusions on the effects of thermal conduction on
the shocked cloud and on the applicability of the diagnostics devised here.

In our model, we do not account for the possible effect of the
back-reaction of accelerated cosmic rays on shock dynamics. In the case
of high Mach number shocks, a part of the shock power may be dissipated
into cosmic rays acceleration, resulting in the increase of the shock
compression ratio. The distribution function of non-thermal particles and
the bulk flow profile in the shock upstream region are sensitive to the
total compression ratio. Thus, even a moderate efficiency of particle
acceleration may reduce the post-shock ion and electron temperatures
(see, e.g. Eq. 18 in \citealt{2008SSRv..134..119B}), with implications
on the X-ray emission. This effect is expected to be large for shocks
with high Mach number (as, for instance, in young SNRs), but not in
middle-aged SNRs (to which this paper is focussed on) for which no
non-thermal emission has been detected. Even if particle acceleration
were not negligible, the effect relevant to our diagnostics would only
be to slightly reduce the efficiency of the thermal conduction because
of the lower post-shock temperature.  As in the case of magnetized clouds
(\citealt{2008ApJ...678..274O}), shocked clouds with considerable particle
acceleration would fall in between the limit of completely suppressed
thermal conduction (HY runs) and the unmagnetized limit with conduction
(RC runs) discussed in this paper.

Recently \cite{2009MNRAS.394.1351P} have demonstrated that the
turbulence plays an important role in shock–cloud interactions, and
that environmental turbulence adds a new dimension to the parameter
space. In particular these authors have shown that the turbulence
is mainly generated around the cloud boundary and in the cloud wake
after $\sim \tau_{\rm cc}$; the main effect is that clouds subject to
a highly turbulent post-shock environment are destroyed significantly
quicker than those within a smooth flow: the larger the cloud density
contrast $\chi$, the higher is the effect of turbulence (for instance,
for $\chi \approx 100$, the effect of the post-shock turbulence dominates
in the shock-cloud interaction). On the other hand, an efficient thermal
conduction makes the cloud boundary smooth very quickly (see Paper I),
and the turbulence grows more slowly around clouds with a smooth density
profile (\citealt{2009MNRAS.394.1351P}). Thus we are confident that our
results are valid and the diagnostics proposed here can be applied to
the clouds with moderate density contrast (i.e. the effects of turbulence
poorly influence the shock-cloud interaction at early evolutionary stages
for $t< \tau_{\rm cc}$) and smooth density profiles (i.e. the growth of
turbulence around clouds is slow), that we considered here.

Finally, our model does not account for the incomplete electron-ion
temperature equilibration in the post-shock region. Equilibrium may not be
complete early during the shock-cloud interaction ($t\approx 0.1~\tau_{\rm
cc}$). In that phase emission models including non-equilibrium should
be applied and the initial part of the light-curves presented in
Fig.~\ref{fig2} may change. On the other hand, the diagnostics discussed
in this section refer to the shock-cloud interaction at $0.4~ \tau_{\rm
cc} \leq t \leq 1.4~ \tau_{\rm cc}$, when the hypothesis of temperature
equilibration can be considered realistic for the shock velocities
explored here (\citealt{2003ApJ...590..846R}).}

%________________________________________________________________
\section{Summary and conclusion}
\label{conc}

In a series of previous paper (Paper I, II,
\citealt{2008ApJ...678..274O}), we have investigated the X-ray emission
arising from the interaction of SNR shock waves with isolated gas clouds
with the scope of identifying the plasma structures that mainly contribute
to X-ray emission detectable with current X-ray instruments. In this work,
we extend our parameter space considering clouds with different value
of density contrast and devise diagnostics in the X-ray band revealing
the shock speed, which is one of the fundamental parameter governing the
shock-cloud interactions, and the cloud evaporation under the effect of
the thermal conduction.

In particular, by performing a series of spectral fittings to the
simulated data of the shock-cloud interaction region, we proved that
there are at least two interesting diagnostic diagrams that can be used:

\begin{itemize}

\item the median energy vs. count rate scatter plot (Fig. \ref{fig6}),
which gives a direct estimate of the shock speed and a hints about the
effects of the thermal conduction;

\item the temperature and emission measure ratios between the knot and
diffuse region of the cloud (Fig. \ref{fig7}), which gives a direct
estimate of the role of the thermal conduction in the evolution of
the system.

\end{itemize}

We stress that the regions (the knot and the DR) which must be selected
to derive the diagnostic diagrams are defined entirely on the basis
of mean photon energy maps, and not on the actual shape of the X-ray
emission.  Therefore, the method is very well-posed and independent,
in first approximation, to the actual shape of the ISM clouds, which
could be more complex that the ideal spherical cases considered in
our hydrodynamic simulations. While this method cannot be considered
a substitution for a detailed approach to the study of shock-cloud
interactions based on the developing of ad-hoc (and time-consuming)
numerical models, nevertheless, we have demonstrated that the diagnostic
diagrams we have presented can be very useful to derive some of the
parameters of the system and the role of thermal conduction in a very
quick and straightforward way. These information can be used in turn
for a more detailed model, if necessary.

The method can be applied to imaging X-ray observations of middle-aged
thermal SNR shells (like Vela or Cygnus Loop), as for instance those
obtained by the XMM-Newton and Chandra X-ray satellites. We used
as a benchmark the XMM-Newton/EPIC observations of the Vela FilD
region of \citet{2005A&A...442..513M}, from which, independently,
\cite{2006A&A...458..213M} have found strong evidence of thermal
conduction at work during the shock-cloud interaction, using a detailed
ad-hoc numerical model. We found that our method is quite effective
in recovering quickly the shock speed and the effects of the thermal
conduction.

\bigskip
\acknowledgements{We thank the referee for constructive and helpful
criticism. The software used in this work was in part developed by
the DOE-supported ASC / Alliance Center for Astrophysical Thermonuclear
Flashes at the University of Chicago. The simulations have been
executed at CINECA (Bologna, Italy) in the framework of the INAF-CINECA
agreement on ``High Performance Computing resources for Astronomy and
Astrophysics'', and on the SCAN (Sistema di Calcolo per l'Astrofisica
Numerica) HPC facility of the INAF\,-\,Osservatorio Astronomico di
Palermo. This work was supported in part by Ministero dell'Universit\`a
e della Ricerca and by Istituto Nazionale di Astrofisica.}

\bibliographystyle{aa}
\bibliography{references}

\end{document}